\documentclass{ptephy_v1}
\usepackage{bm,latexsym,amssymb,amsmath,mathrsfs}
\usepackage{color}
\usepackage{times,setspace}

\newcommand{\x}{\vec{x}}
\newcommand{\e}{\vec{e}}

\begin{document}
\title{Real-time simulation of (2+1)-dimensional lattice gauge theory on qubits}
\author{Arata Yamamoto}
\affil{Department of Physics, The University of Tokyo, Tokyo 113-0033, Japan}

\begin{abstract}
We study the quantum simulation of $Z_2$ lattice gauge theory in 2+1 dimensions.
The dual variable formulation, the so-called Wegner duality, is utilized for reducing redundant gauge degrees of freedom.
The problem of artificial charge unconservation is resolved for any charge distribution.
As a demonstration, we simulate the real-time evolution of the system with two static electric charges, i.e., with two temporal Wilson lines.
Some results obtained by the simulator (with no hardware noise) and the real device (with sizable hardware noise) of a quantum computer are shown.
\end{abstract}

\subjectindex{B01}

\maketitle

\section{Introduction}

Over the past few decades, lattice gauge theory has revealed many equilibrium properties of quantum field theory.
Now we are entering a new era of lattice gauge theory.
The simulation device is changing from classical computers to quantum computers.
Quantum simulation will provide us novel results which cannot be obtained by classical simulation.
One of the main issues is non-equilibrium or real-time dynamics of quantum field theory.
Up to now, quantum simulation was mainly applied to (1+1)-dimensional gauge theories because of limited numbers of qubits \cite{Banerjee:2012pg,Bazavov:2015kka,Martinez_2016,Muschik:2016tws,Zache:2018jbt,Klco:2018kyo,Zhang:2018ufj,Kokail_2019,Surace_2020,Klco:2019evd,Magnifico:2019kyj,Gustafson:2019vsd,Chakraborty:2020uhf,Kharzeev:2020kgc,Shaw:2020udc}.
The application to higher dimensions is compulsory for the future.
There are proposals for higher-dimensional gauge theories \cite{Zohar:2011cw,Zohar:2012ay,Tagliacozzo:2012vg,Tagliacozzo:2012df,Zohar:2013zla,Marcos:2014lda,Mezzacapo:2015bra,Zohar:2016iic,Zohar:2016wmo,Bender:2018rdp}, but further investigation is required.

As the simplest setup, let us consider the $Z_2$ lattice gauge theory without matter fields in 2+1 dimensions.
This is the simplest but interesting theory which shares many essential features with realistic gauge theories.
In 1+1 dimensions, gauge field dynamics is rather trivial.
Gauge fields can be removed from Hamiltonian by solving the Gauss law constraint.
In 2+1 dimensions, this is impossible.
Redundant degrees of freedom must be removed in nontrivial manners.
If the redundant degrees of freedom remain, a notorious problem, noise-induced gauge symmetry violation, can arise \cite{Yang:2020yer}.
It is of great importance to develop redundancy-free or gauge-invariant formulation \cite{kaplan2018gausss,Unmuth-Yockey:2018xak,Stryker:2018efp,Raychowdhury:2018osk,Halimeh:2020xfd,Lamm:2020jwv,haase2020resource,Halimeh:2020ecg,bender2020gauge,1812755}.
In the (2+1)-dimensional $Z_2$ lattice gauge theory, there already exists well-established formulation, say, the Wegner duality \cite{Wegner}.
The formulation can be utilized for quantum simulation, as demonstrated in this paper.

In the classical simulations of pure lattice gauge theory, the most frequently-computed observable is the Wilson line or the Wilson loop.
The Wilson line is interpreted as the world line of an electric charge.
At equilibrium, it is an order parameter for the phase of gauge theory.
In real-time simulation, the expectation value of the Wilson line itself would not be so important because it is no longer an order parameter.
Rather, the response of the system to the Wilson lines would be interesting.
It is interpreted as gauge field dynamics induced by electric charges.
In this paper, we discuss how to perform the real-time simulation of pure lattice gauge theory with electric charges by quantum computers.

\section{$Z_2$ lattice gauge theory}

Let us start with the basics of $Z_2$ lattice gauge theory.
We consider the two-dimensional square lattice in the $x$-$y$ plane.
The $Z_2$ gauge fields are defined on links and their quantum operators are given by the Pauli matrices.
The familiar form of the Hamiltonian \cite{Kogut:1979wt} is
\begin{equation}
\label{eqHoriginal}
 H = - \sum_{\x} \sum_{j=1,2} \sigma_1(\x,j) - \lambda \sum_{\x} \sigma_3(\x,1) \sigma_3(\x,2) \sigma_3(\x+\e_{2},1) \sigma_3(\x+\e_{1},2) 
,
\end{equation}
where $\e_{1}$ and $\e_{2}$ are the unit vector in the $x$ and $y$ directions, respectively.
The first term is the contribution of the electric field, which is defined on links, and the second term is the contribution of the magnetic field, which is defined on plaquettes.
The operator
\begin{equation}
\label{eqGauss}
 G(\x) = \sigma_1(\x,1) \sigma_1(\x-\e_{1},1)\sigma_1(\x,2) \sigma_1(\x-\e_{2},2)
\end{equation}
satisfies the lattice version of the Gauss law
\begin{equation}
\label{eqGauss}
  G(\x) = (-1)^{Q(\x)}
\end{equation}
for physical states.
The charge distribution $Q(\x)$ is set to 0 when an electric charge does not exist at $\x$ and to 1 when an electric charge exists at $\x$.
Although we could in principle consider time-dependent $Q(\x)$, i.e., moving charges, we only consider time-independent $Q(\x)$ in this paper.
Because of the commutation relation
\begin{equation}
\label{eqCC}
 [H,G(\x)] = 0 
,
\end{equation}
$Q(\x)$ is conserved at each $\x$.

It is easy to put static electric charges on the system.
What we need to do is just to prepare an initial state with nonzero charge distribution.
The charge distribution does not change in time evolution because of Eq.~\eqref{eqCC}.
Thus, static electric charges would be realized in ideal simulation.
In quantum computers, however, charge conservation is artificially violated due to device noises.
The charge distribution cannot be kept fixed.
It is essential to use noise-robust formulation to realize static electric charges.

When the lattice size is $L_x \times L_y$, the number of links is $\sim 2L_xL_y$.
(The symbol ``$\sim$'' means that the precise number depends on boundary conditions.)
The dimension of the total Hilbert space is $\sim 2^{2L_xL_y}$, but we do not need to treat the total Hilbert space.
The total Hilbert space is divided into $\sim 2^{L_xL_y}$ subspaces with different distribution of $Q(\x)$.
Since the subspaces are decoupled with each other, we only have to treat one subspace in one simulation.
Removing unnecessary subspaces, we can reduce computational cost and suppress artificial process.

\section{Dual variable formulation}

\begin{figure}[h]
\begin{center}
 \includegraphics[width=.8\textwidth]{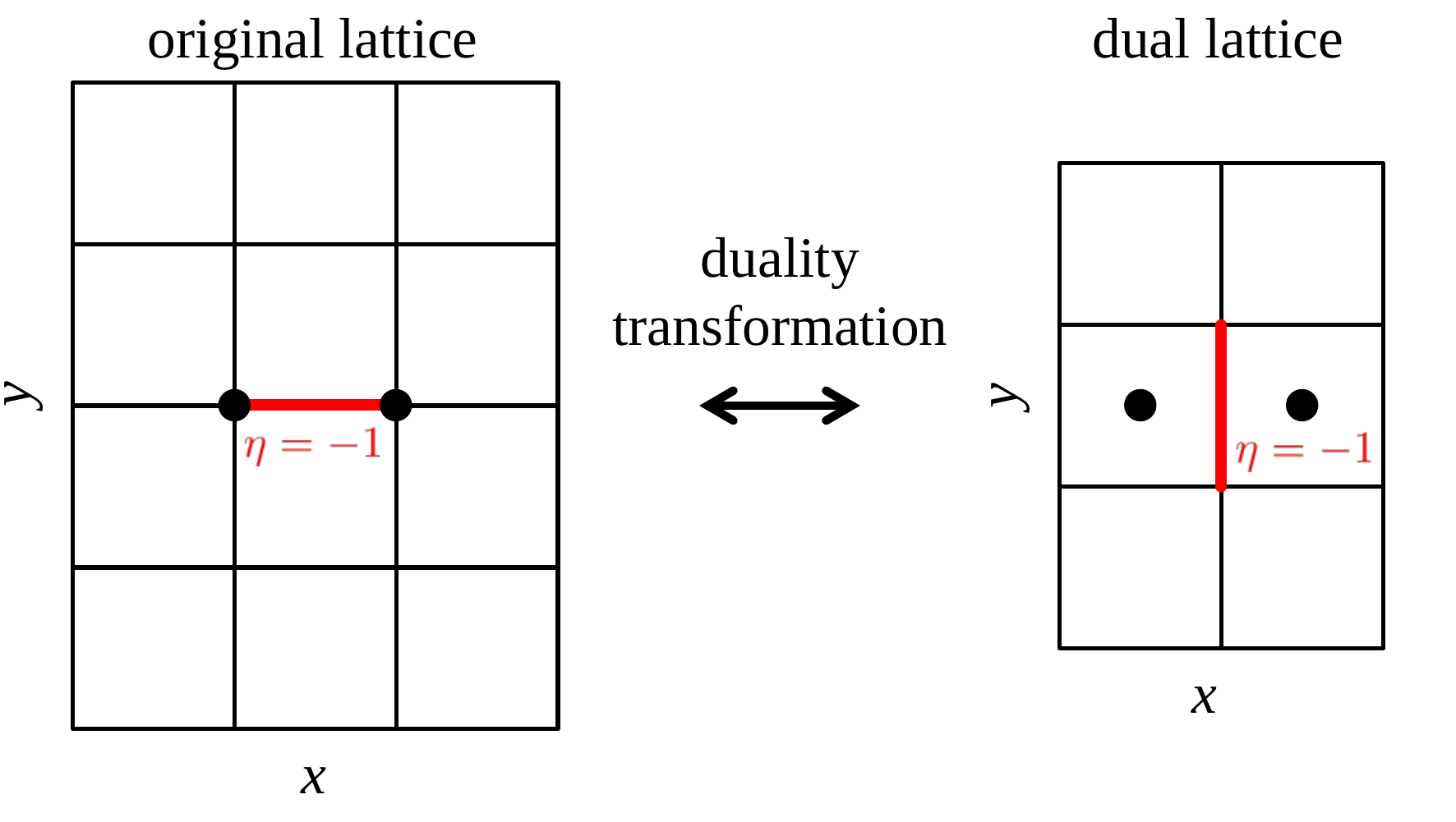}
\caption{
\label{figschematic}
Original lattice and dual lattice.
Two static electric charges are located at the circles.
The phase factors are set to $\eta(\x,1)=\eta(\x^*-\e_{2},2) = -1$ on the red links.
}
\end{center}
\end{figure}

In the (2+1)-dimensional $Z_2$ lattice gauge theory, there is famous formulation to remove redundant degrees of freedom \cite{Wegner}.
The $Z_2$ gauge fields on the original lattice are mapped to the $Z_2$ spin variables on the dual lattice.
The relation between the original lattice and the dual lattice is depicted in Fig.~\ref{figschematic}.
The center of the plaquette on the original lattice defines to the site on the dual lattice.
In the following equations, quantities on the dual lattice are written with an asterisk ``*''.
For example, $\x^*$ denotes the position of a dual site.

When no electric charge exists at all, $Q(x)=0 $ for $\forall x$, the duality transformation is given by two equations \cite{Kogut:1979wt}:
the dual spin-flip operator
\begin{equation}
\label{eqdual1}
\sigma_1(\x^*) = \sigma_3(\x,1) \sigma_3(\x,2) \sigma_3(\x+\e_{2},1) \sigma_3(\x+\e_{1},2)
\end{equation}
and the dual spin operator
\begin{equation}
\label{eqdual2}
 \sigma_3(\x^*) = \prod_{n\ge 0} \sigma_1(\x-n\e_2,1)
.
\end{equation}
Boundary conditions are assumed to be non-periodic.
When electric charges exist, the second equation must be modified to satisfy the Gauss law.
Let us define the phase factor $\eta(\x,j)$ such that $\eta(\x,j) = -1$ for the links connecting the electric charges and $\eta(\x,j) = 1$ elsewhere (see Fig.~\ref{figschematic}).
The phase factor satisfies the equation
\begin{equation}
\label{eqeta}
\eta(\x,1) \eta(\x-\e_{1},1) \eta(\x,2) \eta(\x-\e_{2},2) = (-1)^{Q(x)}
\end{equation}
by definition.
(The choice for the path connecting the electric charges is not unique.
The path ambiguity is equivalent to the redefinition of field variables.
Even if the path is deformed, Eq.~\eqref{eqeta} holds and physical results are invariant.)
The duality transformation \eqref{eqdual2} is generalized as
\begin{equation}
\label{eqdual3}
\sigma_3(\x^*)=\prod_{n\ge 0} \eta(\x-n\e_2,1) \sigma_1(\x-n\e_2,1)
.
\end{equation}
From Eq.~\eqref{eqdual3} and the Gauss law constraint \eqref{eqGauss}, we get
\begin{eqnarray}
 \sigma_1(\x,1) &=& \eta(\x^*-\e_{2},2) \sigma_3(\x^*-\e_{2}) \sigma_3(\x^*)
\\
 \sigma_1(\x,2) &=& \eta(\x^*-\e_{1},1) \sigma_3(\x^*-\e_{1}) \sigma_3(\x^*)
\end{eqnarray}
under non-periodic boundary conditions.
The dual phase factor is defined by the original phase factor crossing the dual link, $\eta(\x^*-\e_{1},1) = \eta(\x,2)$ and $\eta(\x^*-\e_{2},2) = \eta(\x,1)$.
After all, the dual Hamiltonian
\begin{equation}
\label{eqHdual}
 H = - \sum_{\x^*} \sum_{j=1,2} \eta(\x^*,j) \sigma_3(\x^*) \sigma_3(\x^*+\e_{j}) - \lambda \sum_{\x^*} \sigma_1(\x^*) 
\end{equation}
is obtained.

The dual Hamiltonian \eqref{eqHdual} has three advantages compared with the original Hamiltonian \eqref{eqHoriginal}.
First, the Gauss law constraint is automatically satisfied.
It is easy to probe
\begin{equation}
\begin{split}
 G(\x) 
&= \sigma_1(\x,1) \sigma_1(\x-\e_{1},1)\sigma_1(\x,2) \sigma_1(\x-\e_{2},2)
\\
&= \{ \eta(\x^*-\e_{2},2) \sigma_3(\x^*-\e_{2}) \sigma_3(\x^*) \} \{ \eta(\x^*-\e_{1}-\e_{2},2) \sigma_3(\x^*-\e_{1}-\e_{2}) \sigma_3(\x^*-\e_{1}) \}
\\
&\quad \times \{ \eta(\x^*-\e_{1},1) \sigma_3(\x^*-\e_{1}) \sigma_3(\x^*) \} \{ \eta(\x^*-\e_{1}-\e_{2},1) \sigma_3(\x^*-\e_{1}-\e_{2}) \sigma_3(\x^*-\e_{2}) \}
\\
&= \eta(\x^*-\e_{2},2) \eta(\x^*-\e_{1}-\e_{2},2) \eta(\x^*-\e_{1},1) \eta(\x^*-\e_{1}-\e_{2},1)
\\
&= \eta(\x,1) \eta(\x-\e_{1},1) \eta(\x,2) \eta(\x-\e_{2},2)
\\
&= (-1)^{Q(x)}
.
\end{split}
\end{equation}
This is an identity equation, so charge conservation is exact.
Second, required memory size is smaller.
While the number of the gauge fields is $\sim 2L_xL_y$ in the original Hamiltonian, the number of the dual spin variables is $\sim L_xL_y$ in the dual Hamiltonian.
Third, the implementation on quantum gates is easier.
The original Hamiltonian includes the product of four Pauli matrices.
It will be implemented by some complicated combination of multi-qubit operations.
The dual Hamiltonian only includes the product of two Pauli matrices.
It can be directly implemented by a two-qubit operation.

\section{Real-time simulation}

The time evolution of the total system is given by
\begin{equation}
|\Psi(t)\rangle = e^{-iHt} |\Psi(0)\rangle
.
\end{equation}
The continuous time evolution is approximated by the $n$-times matrix operation via the Suzuki-Trotter decomposition,
\begin{equation}
e^{-iHt} \simeq  \left( \prod_{x^*} e^{-i \mathcal{H}_{Ex} \delta t} \prod_{x^*} e^{-i \mathcal{H}_{Ey} \delta t} \prod_{x^*} e^{-i\mathcal{H}_B \delta t} \right)^n
\end{equation}
with
\begin{eqnarray}
 \mathcal{H}_{Ex} &=& - \eta(x^*-\e_{2},2) \sigma_3(x^*-\e_{2}) \sigma_3(x^*)
\\
 \mathcal{H}_{Ey} &=& - \eta(x^*-\e_{1},1) \sigma_3(x^*-\e_{1}) \sigma_3(x^*)
\\
 \mathcal{H}_B &=& - \lambda \sigma_1(x^*)
.
\end{eqnarray}
The time step $\delta t$ must be small enough to justify the Suzuki-Trotter approximation.
Each matrix operation can be easily implemented by quantum gates.
The matrix $e^{-i \mathcal{H}_E \delta t}$ is realized by the controlled rotation gate $C_{R_z}$ for two qubits and the standard rotation gate $R_z$ for a single qubit.
The matrix $e^{-i\mathcal{H}_B \delta t}$ is realized by the standard rotation gate $R_x$ for a single qubit.

\begin{figure}[p]
\begin{center}
\begin{minipage}{0.4\textwidth}
\begin{center}
 \includegraphics[width=1\textwidth]{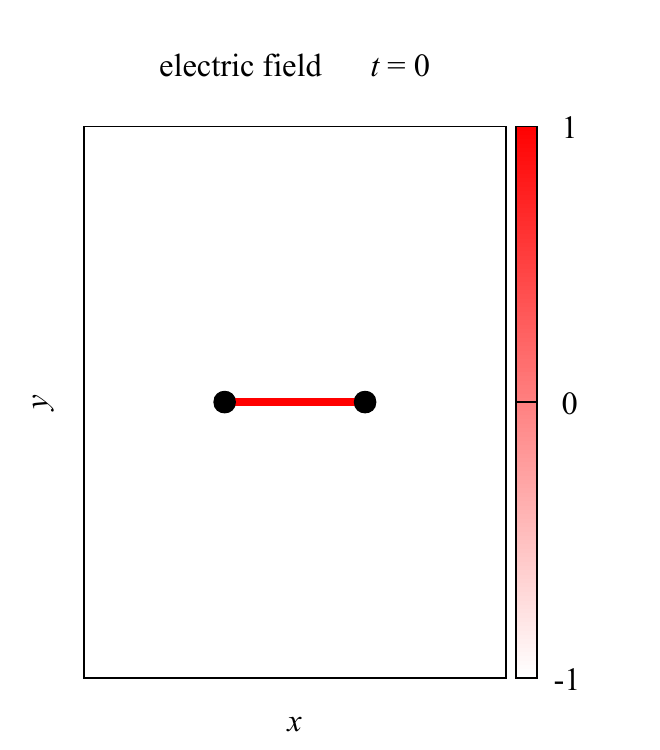}
\end{center}
\end{minipage}
\begin{minipage}{0.4\textwidth}
\begin{center}
 \includegraphics[width=1\textwidth]{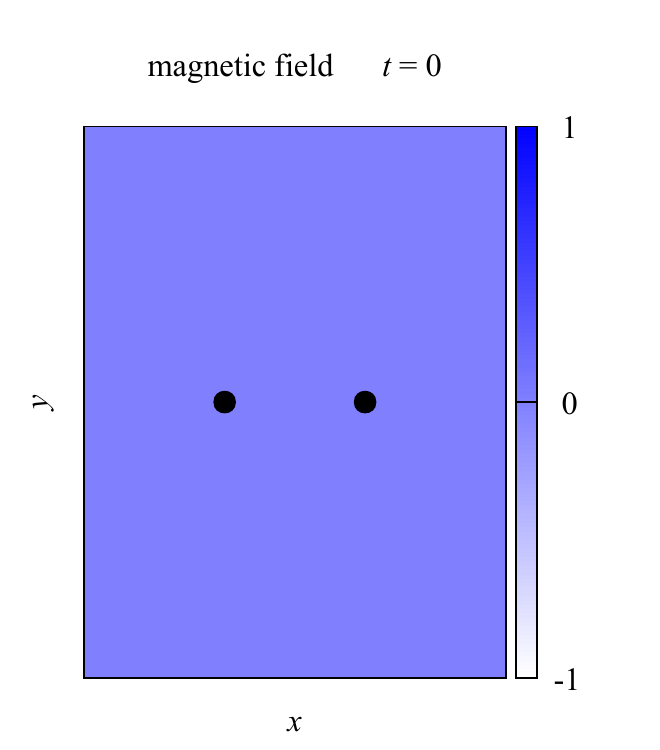}
\end{center}
\end{minipage}
\\
\begin{minipage}{0.4\textwidth}
\begin{center}
 \includegraphics[width=1\textwidth]{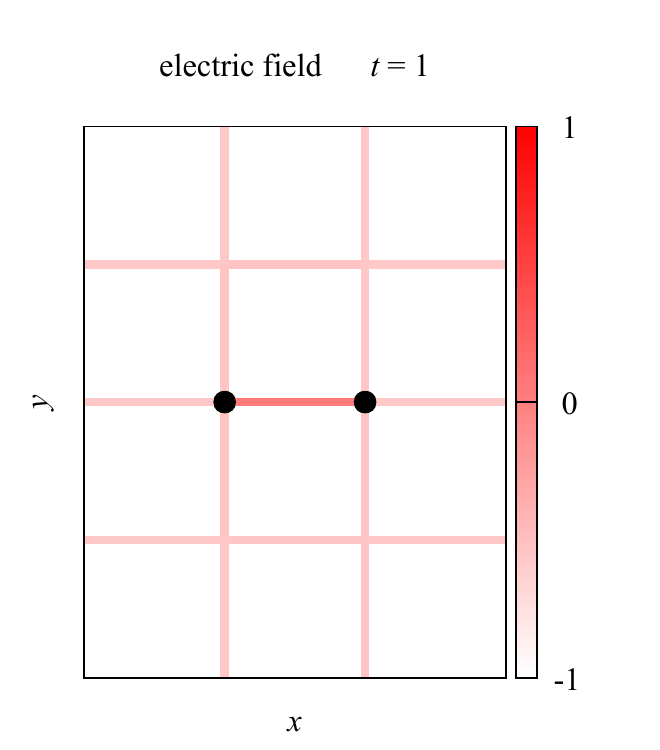}
\end{center}
\end{minipage}
\begin{minipage}{0.4\textwidth}
\begin{center}
 \includegraphics[width=1\textwidth]{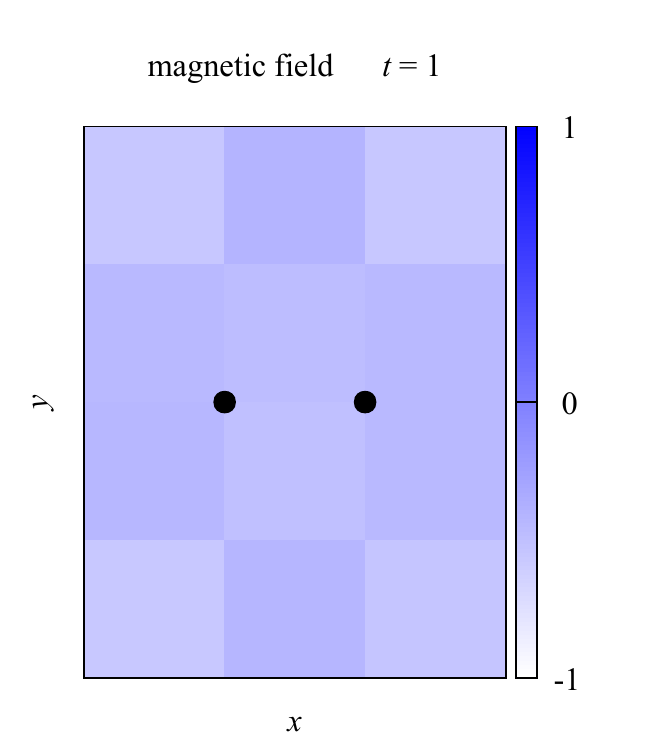}
\end{center}
\end{minipage}
\begin{minipage}{0.4\textwidth}
\begin{center}
 \includegraphics[width=1\textwidth]{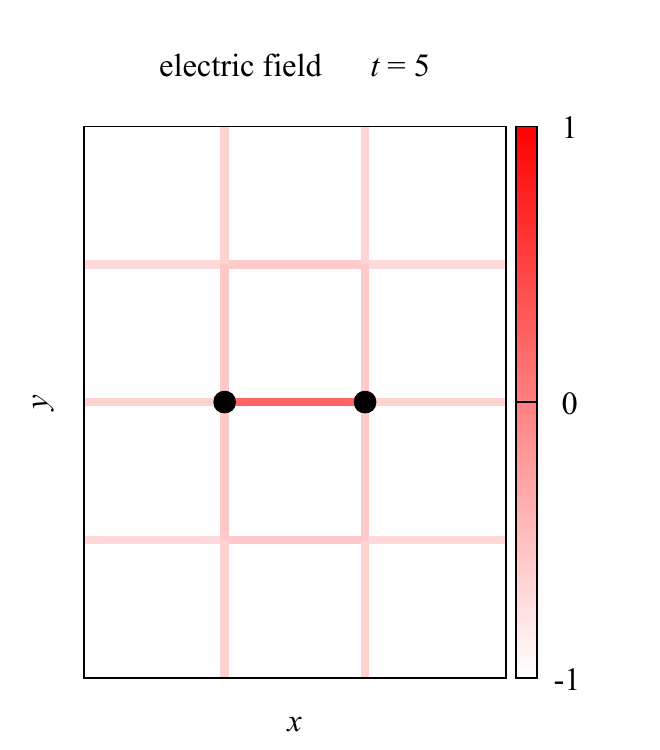}
\end{center}
\end{minipage}
\begin{minipage}{0.4\textwidth}
\begin{center}
 \includegraphics[width=1\textwidth]{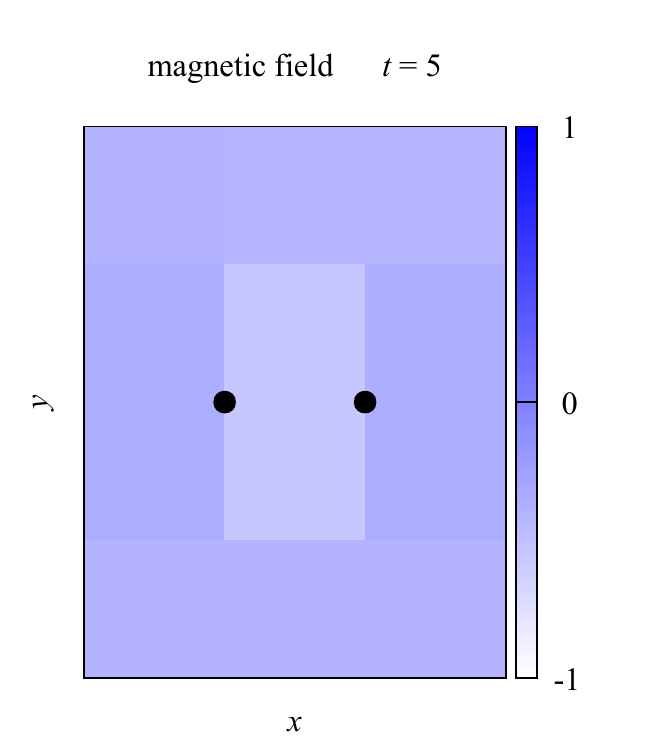}
\end{center}
\end{minipage}
 \caption{
\label{figEB}
Energy distribution at $t=0$ (top), $t=1$ (middle), and $t=5$ (bottom).
The electric field energy $\langle \Psi | (\mathcal{H}_{Ex} + \mathcal{H}_{Ey}) |\Psi \rangle$ is defined on links (left) and the magnetic field energy $\langle \Psi | \mathcal{H}_{B} |\Psi \rangle$ is defined on plaquettes (right).
The results are obtained by the simulator with the time step $\delta t =0.10$.
The statistical errors are omitted.
}
\end{center}
\end{figure}

We computed the time evolution by the simulator and the real device of a quantum computer.
The simulator is the algorithm on a classical computer designed to mimic a quantum computer.
For the real device, we used ``ibmq\_16\_melbourne'', which has 15 qubits, in IBM Quantum services \cite{IBM}.
The computation was performed with the dual Hamiltonian on the dual lattice, and then the obtained results were translated to the language on the original lattice.
The geometry of the lattice is shown in Fig.~\ref{figschematic}.
The dual lattice is the $3\times4$ lattice with open boundary conditions.
Two static electric charges are located at $\x=(x,y)=(1,2)$ and $(2,2)$.
Since the dual spin variable is given by two quantum states $|1\rangle$ and $|-1\rangle$, it can be embedded into a digital qubit.
The initial state is set as
\begin{equation}
\label{eqinitial}
 |\Psi(0)\rangle = \prod_{x^*} |1\rangle
.
\end{equation}
The parameter is fixed at $\lambda=1$.

The electric field energy $\langle \Psi | (\mathcal{H}_{Ex} + \mathcal{H}_{Ey}) |\Psi \rangle$ is defined on links.
It can be measured by counting the probability of each state of $|\Psi\rangle$ because the matrix $\sigma_3(x^*-\e_{j}) \sigma_3(x^*)$ is diagonal.
The magnetic field energy $\langle \Psi | \mathcal{H}_{B} |\Psi \rangle$ is defined on plaquettes.
It can be measured by diagonalizing the matrix as $\sigma_1 = h^\dagger \sigma_3 h$  with the Hadamard gate $h$.
The distributions of these energies are shown in Fig.~\ref{figEB}.
At the initial time $t=0$, the distributions can be analytically calculated from the initial state \eqref{eqinitial}.
The electric field energy is $1$ on the link between the electric charges and $-1$ on all the other links.
The magnetic field energy is $0$ on all the plaquettes.
After time evolution, the electric field energy spreads out all over the lattice and the magnetic field energy is transfered to the electric field energy. 
The snapshots at $t=1$ and $t=5$ are shown in the figure.

\begin{figure}[h]
\begin{minipage}{0.5\textwidth}
\begin{center}
 \includegraphics[width=1\textwidth]{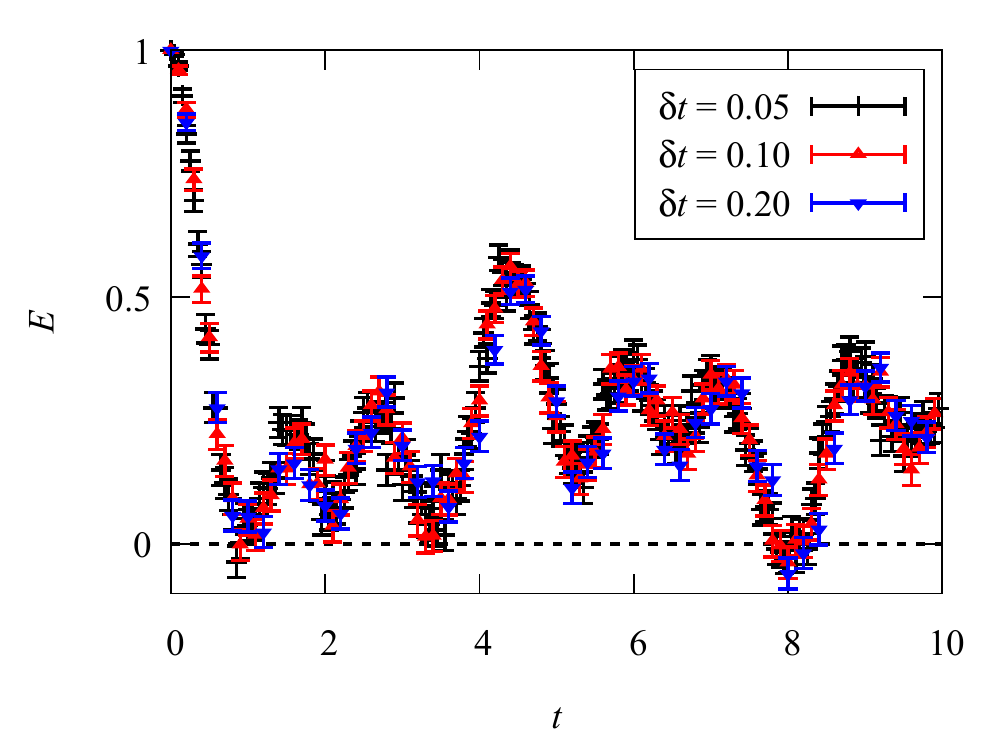}
\end{center}
\end{minipage}
\begin{minipage}{0.5\textwidth}
\begin{center}
 \includegraphics[width=1\textwidth]{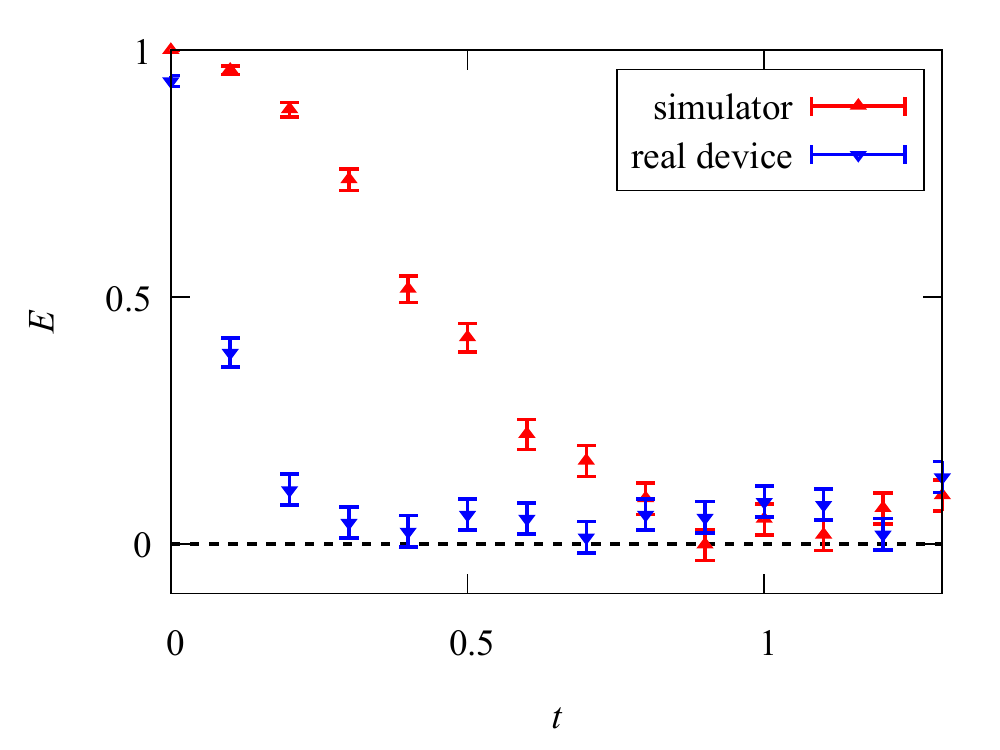}
\end{center}
\end{minipage}
\caption{
\label{figdt}
Dependence on the time step $\delta t$ (left) and the difference between the simulator and the real device of a quantum computer (right).
The electric field energy on the link between the electric charges, $E \equiv \langle \Psi | \mathcal{H}_{Ex} |\Psi \rangle$, is plotted.
The error bars are statistical errors.
}
\end{figure}

Two systematic errors are analyzed in Fig.~\ref{figdt}.
The electric field energy on the link between the electric charges is plotted as one typical observable.
In the left panel, we show the time-step dependence.
The results obtained by the simulator with $\delta t=0.05,0.10$ and $0.20$ are shown.
They show good agreement.
This indicates that the Suzuki-Trotter error is sufficiently small.
In the right panel, we compare the result obtained by the simulator and the raw data obtained by the real device (without any error mitigation).
The simulator is classical simulation, so the result is the exact answer without noise.
On the other hand, the real device suffers from many kinds of hardware noise, e.g., gate errors, readout errors, etc \cite{Gustafson:2019vsd,Funcke:2020olv}.
Furthermore, in the real device, the geometry of qubits is different from the geometry of the simulated lattice.
The two-qubit operation is reconstructed as a sequence of the operations, and thus leads to noise enhancement.
The raw data cannot reproduce the simulator results.
Although the artificial violation of charge conservation is absent, the deviation is still large.
Error mitigation will be necessary for practical use.

\section{Summary and outlook}

We have studied the real-time evolution of $Z_2$ lattice gauge theory with static electric charges.
The Wegner duality was originally discovered in the $Z_2$ gauge theory \cite{Wegner}.
Later, the dual variable formulation was generalized to gauge theories with $Z_N$ subgroups \cite{Ukawa:1979yv}.
For example, it is applicable to SU($N$) gauge theory.
The implementation of SU($N$) gauge theory is now difficult because of the limitation of computational resource.
More qubits are required for digitization of continuous gauge groups \cite{Alexandru:2019nsa,Ji:2020kjk}.
Although it might be a long journey, it will become possible to analyze real-time dynamics of gluons around color charges, such as the time evolution of a confining string, someday in the future.

\ack
The author was supported by JSPS KAKENHI Grant No.~19K03841.   
The author acknowledges the use of IBM Quantum services for this work.
The views expressed are those of the author, and do not reflect the official policy or position of IBM or the IBM Quantum team.

\bibliographystyle{ptephy}
\bibliography{paper}

\end{document}